\documentclass[a4paper,11pt,english]{article}
\usepackage{fixltx2e} \usepackage{cmap} \usepackage{babel}
\usepackage[T1]{fontenc}
\usepackage[latin1]{inputenc}
\usepackage{ifthen}
\usepackage{float} 
\usepackage{graphicx}
\usepackage{xcolor} \usepackage{ulem} \usepackage{changebar}
\newcommand\TLSdel[1]{\cbstart{}\textcolor{blue}{\uline{#1}}\cbend{}}
\newcommand\TLSins[1]{\cbdelete{}\textcolor{red}{\sout{#1}}}
\renewcommand\TLSdel[1]{}
\renewcommand\TLSins[1]{#1}

\usepackage{graphicx}
\usepackage{url}
\usepackage{hyperref}

\usepackage{cise-like}
\usepackage{enumitem}
\setdescription{font=\normalfont\sffamily\bfseries, itemsep=.5ex,
    parsep=.5ex, leftmargin=3ex}
\usepackage{lmodern}

\def\emph#1{{\sl #1}}

\providecommand{\DUadmonition}[2][class-arg]{\ifcsname DUadmonition#1\endcsname\csname DUadmonition#1\endcsname{#2}\else
    \begin{center}
      \fbox{\parbox{0.9\textwidth}{#2}}
    \end{center}
  \fi
}

\providecommand*{\DUrole}[2]{\ifcsname DUrole#1\endcsname\csname DUrole#1\endcsname{#2}\else\ifcsname docutilsrole#1\endcsname\csname docutilsrole#1\endcsname{#2}\else#2\fi\fi}

\newenvironment{DUlegend}{\small}{}
{}

\providecommand*{\DUtitle}[2][class-arg]{\ifcsname DUtitle#1\endcsname\csname DUtitle#1\endcsname{#2}\else
    \smallskip\noindent\textbf{#2}\smallskip\fi
}

\providecommand*{\DUroletitlereference}[1]{\textsl{#1}}

\providecommand*{\DUtransition}[1][class-arg]{\hspace*{\fill}\hrulefill\hspace*{\fill}
  \vskip 0.5\baselineskip
}

\ifthenelse{\isundefined{\hypersetup}}{
  \usepackage[colorlinks=true,linkcolor=blue,urlcolor=blue]{hyperref}
}{}
\hypersetup{
  pdftitle={The NumPy array: a structure for efficient numerical computation},
}

\begin{document}

\title{The NumPy array: a structure for efficient numerical computation\phantomsection\label{the-numpy-array-a-structure-for-efficient-numerical-computation}}
\author{
{\bf\sffamily Stéfan van der Walt}, Stellenbosch University {\small\sc South
Africa} \\
{\bf\sffamily S. Chris Colbert}, Enthought {\small\sc USA}\\
{\bf\sffamily Gael Varoquaux}, INRIA Saclay {\small\sc France}
}
\date{
\begin{minipage}{.9\linewidth}
This article is published in {\sl IEEE
Computing in Science and Engineering.} Please refer to the published
version if accessible, as it contains editor's improvements.
(c) 2011 IEEE.%
\footnote{%
Personal use of this material is permitted. Permission
from IEEE must be obtained for all other users, including reprinting/
republishing this material for advertising or promotional purposes,
creating new collective works for resale or redistribution to servers or
lists, or reuse of any copyrighted components of this work in other
works.}
\end{minipage}
} 
\maketitle

In the Python world, NumPy arrays are the standard representation for
numerical data. Here, we show how these arrays enable efficient
implementation of numerical computations in a high-level language.
Overall, three techniques are applied to improve performance:
vectorizing calculations, avoiding copying data in memory, and
minimizing operation counts.

We first present the NumPy array structure, then show how to use it
for efficient computation, and finally how to share array data with
other libraries.

\section*{Introduction\phantomsection\addcontentsline{toc}{section}{Introduction}\label{introduction}}

The Python programming language provides a rich set of high-level data
structures: lists for enumerating a collection of objects,
dictionaries to build hash tables, etc. However, these structures are
not ideally suited to high-performance numerical computation.

In the mid-90s, an international team of volunteers started to develop
a data-structure for efficient array computation.  This structure
evolved into what is now known as the \protect\TLSdel{NumPy} N-dimensional \protect\TLSins{NumPy} array.

The NumPy package, which comprises the NumPy array as well as a set of
accompanying mathematical functions, has found wide-spread adoption in
academia, national laboratories, and industry, with applications
ranging from gaming to space exploration.

A NumPy array is a multidimensional, uniform collection of \protect\TLSdel{elements
(in other words, all elements occupy the same number of bytes in
memory).} \protect\TLSins{elements.}
An array is characterized by the type of elements it contains and by
its shape.  For example, a matrix may be represented as an array of
shape \DUrole{math}{($M \times N$)} that contains numbers, e.g., floating point
or complex numbers. Unlike matrices, NumPy arrays can have \protect\TLSdel{up to 32 dimensions.} \protect\TLSins{any
dimensionality.} Furthermore, they may contain other kinds of elements
(or even combinations of elements), such as booleans or dates.

Underneath the hood, a NumPy array is really just a convenient way of
describing one or more blocks of computer memory, so that the numbers
represented may be easily manipulated.  \protect\TLSdel{As discussed in this article,
NumPy provides a high-level abstraction for numerical computation,
without compromising performance.}

\subsection*{Basic
usage\phantomsection\addcontentsline{toc}{subsection}{Basic
usage}\label{basic-usage-sidebar}}

Throughout the code examples in the article, we assume that NumPy is
imported as follows:
\begin{verbatim}
import numpy as np
\end{verbatim}

Code snippets are shown as they appear inside an
[\hyperlink{ipython}{IPython}] prompt, such as this:

\begin{verbatim}
In  [3]: np.__version__
Out [3]: '1.4.1'
\end{verbatim}

\protect\TLSdel{IPython is an interactive front-end to Python with TAB-completion and
easy documentation inspection.  Lines of the form \texttt{In {[}n{]}:}
represent the input prompt, equivalent to vanilla Python's \texttt{>{}>{}>}.}

Elements contained in an array can be indexed using the \texttt{{[}{]}}
operator. In addition, parts of an array may be retrieved using
standard Python slicing of the form \texttt{start:stop:step}. For instance,
the first two rows of an array \texttt{x} are given by \texttt{x{[}:2, :{]}} or
columns 1 through 3 by \texttt{x{[}:, 1:4{]}}.  Similarly, every second row is
given by \texttt{x{[}::2, :{]}}.  Note that Python uses zero-based \protect\TLSins{indexing.}\protect\TLSdel{indexing,
and that the slice \texttt{1:3} excludes the last element, 3.
}

\protect\TLSdel{For more information on the Python language, refer to
\href{http://docs.python.org}{http://docs.python.org}.}

\section*{The structure of a NumPy array: a view on memory\phantomsection\addcontentsline{toc}{section}{The structure of a NumPy array: a view on memory}\label{the-structure-of-a-numpy-array-a-view-on-memory}}

A NumPy array (also called an ``ndarray'', short for N-dimensional
array) describes memory, using the following attributes:
\begin{description}

\item[Data pointer] the memory address of the first byte in the array.

\item[Data type description] the kind of elements contained in the
array, for example floating point numbers or integers.

\item[Shape] the shape of the array, for example \texttt{(10, 10)} for a
ten-by-ten array, or \texttt{(5, 5, 5)} for a block of data describing
a mesh grid of x-, y- and z-coordinates.

\item[Strides] the number of bytes to skip in memory to proceed to the
next \protect\TLSdel{element along a given dimension.}
\protect\TLSins{element.}  For a \texttt{(10, 10)} array of bytes, for
example, the strides may be \texttt{(10, 1)}, in other words: proceed one
byte to get to the next column and ten bytes to locate the next row.

\item[Flags] which define whether we are allowed to modify the array,
whether memory layout is C- or Fortran-contiguous\footnote{In C, memory
is laid out in ``row major'' order, i.e., rows are stored one after
another in memory.  In Fortran, \emph{columns} are stored successively.
}, and so forth. 

\end{description}

NumPy's strided memory model deserves particular attention, as it
provides a powerful way of viewing the same memory in different ways
without copying data.  For example, consider the following integer
array:
\begin{verbatim}
# Generate the integers from zero to eight and 
# re pack them into a 3x3 array

In [1]: x = np.arange(9).reshape((3, 3))

In [2]: x
Out[2]:
array([[0, 1, 2],
       [3, 4, 5],
       [6, 7, 8]])
 
In [3]: x.strides
Out[3]: (24, 8)
\end{verbatim}

\medskip
On our 64-bit system, the default integer data-type occupies 64-bits,
or 8 bytes, in memory. The strides therefore describe skipping 3
integers in memory to get to the next row and one to get to the next
column.  We can now generate a view on the same memory where we only
examine every second element.
\begin{verbatim}
In [4]: y = x[::2, ::2]
 
In [5]: y
Out[5]:
array([[0, 2],
       [6, 8]])
 
In [6]: y.strides
Out[6]: (48, 16)
\end{verbatim}

The arrays \texttt{x} and \texttt{y} point to the same memory (i.e., if we
modify the values in \texttt{y} we also modify those in \texttt{x}), but the
strides for \texttt{y} have been changed so that only every second element
is seen along either axis. \texttt{y} is said to be a \emph{view} on \texttt{x}:
\begin{verbatim}
In [7]: y[0, 0] = 100
 
In [8]: x
Out[8]:
array([[100,   1,   2],
       [  3,   4,   5],
       [  6,   7,   8]])
\end{verbatim}

Views need not be created using slicing only.  By modifying strides,
for example, an array can be transposed or reshaped at zero cost (no
memory needs to be copied).  Moreover, the strides, shape and dtype
attributes of an array may be specified manually by the user (provided
they are all compatible), enabling a plethora of ways in which to
interpret the underlying data.
\begin{verbatim}
# Transpose the array, using the shorthand "T"
In [9]: xT = x.T
 
In [10]: xT
Out[10]:
array([[100, 3, 6],
       [  1, 4, 7],
       [  2, 5, 8]])
 
In [11]: xT.strides
Out[11]: (8, 24)
 
# Change the shape of the array
In [12]: z = x.reshape((1, 9))
In [13]: z
Out[13]: array([[100, 1, 2, 3, 4, 5, 6, 7, 8]])
 
In [14]: z.strides
Out[14]: (72, 8)
 
# i.e., for the two-dimensional z, 9 * 8 bytes 
# to skip over a row of 9 uint8 elements,
# 8 bytes to skip a single element
\end{verbatim}
 
\begin{verbatim}
# View data as bytes in memory rather than
# 64bit integers
In [15]: z.dtype = np.dtype('uint8')

In [16]: z
Out[17]:
array([[100, 0, 0, ...,
        0, 0, 0, 0, 0, 0]], dtype=uint8)

In [18]: z.shape
Out[19]: (1, 72)

In [20]: z.strides
Out[20]: (72, 1)
\end{verbatim}

In each of these cases, the resulting array points to the same
memory. The difference lies in the way the data is interpreted, based
on shape, strides and data-type.  Since no data is copied in memory, 
these operations are extremely efficient.

\section*{Numerical operations on arrays: vectorization \protect\TLSdel{and broadcasting}\phantomsection\addcontentsline{toc}{section}{operations on arrays: vectorization and broadcasting\label{numerical-operations-on-arrays-vectorization-and-broadcasting}} }

In any scripting language, unjudicious use of for-loops may lead to
poor performance, particularly in the case where a simple computation
is applied to each element of a large data-set.

Grouping these element-wise operations together, a process known as
vectorisation, allows NumPy to perform such computations much more
rapidly.

Suppose we have a vector \texttt{a} and wish to multiply its magnitude by
3.  A traditional for-loop approach would look as follows:
\begin{verbatim}
In [21]: a = [1, 3, 5]
 
In [22]: b = [3*x for x in a]
 
In [23]: b
Out[23]: [3, 9, 15]
\end{verbatim}

The vectorized approach applies this operation to all elements of an array:

\begin{verbatim}
In [24]: a = np.array([1, 3, 5])
 
In [25]: b = 3 * a
 
In [26]: b
Out[26]: array([ 3,  9, 15])
\end{verbatim}

Vectorized operations in NumPy are implemented in C, resulting in a
significant speed improvement.

Operations are not restricted to interactions between scalars and
arrays.  For example, here NumPy performs a fast element-wise subtraction
of two arrays:

\begin{verbatim}
In  [27]: b - a
Out [27]: array([2,  6, 10])
\end{verbatim}

When the shapes of the two arguments are not the same, but share a
common shape dimension, the operation is \emph{broadcast} across the array.
In other words, NumPy expands the arrays such that the operation
becomes viable:
\begin{verbatim}
In [28]: m = np.arange(6).reshape((2,3))

In  [29]:m
Out [29]:
array([[0, 1, 2],
       [3, 4, 5]])
 
In [30]: b + m
Out[30]:
array([[ 3, 10, 17],
       [ 6, 13, 20]])
\end{verbatim}

Refer to ``Broadcasting Rules'' to see when these operations
are viable.

To save memory, the broadcasted arrays are never physically constructed;
NumPy simply uses the appropriate array elements during
computation\footnote{ Under the hood, this is achieved by using strides
of zero.}

\section*{Broadcasting
Rules\phantomsection\addcontentsline{toc}{section}{Broadcasting
Rules}\label{broadcasting-rules-sidebar}}

Before broadcasting two arrays, NumPy verifies that all dimensions are
suitably matched.  Dimensions match when they are equal, or when
either is 1 or None.  In the latter case, the dimension of the output
array is expanded to the larger of the two.

For example, consider arrays \texttt{x} and \texttt{y} with shapes \texttt{(2, 4, 3)}
and \texttt{(4, 1)} respectively.  These arrays are to be combined in a
broadcasting operation such as \texttt{z = x + y}.  We match their
dimensions as follows:
\begin{center}%
\begin{tabular}{clcr}
x &(2, &4, &3)\\
y &(   &4, &1)\\
\hline
z &(2, &4, &3)\\
\end{tabular}%
\end{center}

Therefore, the dimensions of these arrays are compatible, and yield
and output of shape \texttt{(2, 4, 3)}.

\section*{Vectorization and broadcasting examples\phantomsection\addcontentsline{toc}{section}{Vectorization and broadcasting examples}\label{vectorization-and-broadcasting-examples}}

\subsection*{Evaluating Functions\phantomsection\addcontentsline{toc}{subsection}{Evaluating Functions}\label{evaluating-functions}}

Suppose we wish to evaluate a function \texttt{f} over a large set of
numbers, \texttt{x}, stored as an array.  Using a for-loop, the result is
produced as follows:
\begin{verbatim}
In [31]: def f(x):
    ...:     return x**2 - 3*x + 4
    ...:
 
In [32]: x = np.arange(1e5)
 
In [33]: y = [f(i) for i in x]
\end{verbatim}

On our machine, this loop executes in approximately 500 miliseconds.
Applying the function \texttt{f} on the NumPy array \texttt{x} engages the fast,
vectorized loop, which operates on each element individually:

\begin{verbatim}
In [34]: y = f(x)
 
In [35]: y
Out[35]:
array([ 4.0000e+0,  2.0000e+0,  2.0000e+0, ...,
        9.9991e+9,  9.9993e+9,  9.9995e+9])
\end{verbatim}

The vectorized computation executes in 1 milisecond.

As the length of the input array \texttt{x} grows, however, execution speed
decreases due to the construction of large temporary arrays.  For
example, the operation above roughly translates to
\begin{verbatim}
a = x**2
b = 3*x
c = a - b
fx = c + 4
\end{verbatim}

Most array-based systems do not provide a way to circumvent the
creation of these temporaries.  With NumPy, the user may choose to
perform operations ``in-place''---in other words, in such a way that no
new memory is allocated and all results are stored in the current
array.  
\begin{verbatim}
def g(x):
  # Allocate the output array with x-squared
  fx = x**2
  # In-place operations: no new memory allocated
  fx -= 3*x
  fx += 4
  return fx
\end{verbatim}

Applying \texttt{g} to \texttt{x} takes 600 microseconds; almost twice as fast
as the naive vectorization. Note that we did not compute \texttt{3*x}
in-place, as it would modify the original data in \texttt{x}.

This example illustrates the ease with which NumPy handles vectorized
array operations, without relinquishing control over
performance-critical aspects such as memory allocation.

Note that performance may be boosted even further by using tools such as
[\hyperlink{cython}{Cython}], [\hyperlink{theano}{Theano}] or
[\hyperlink{numexpr}{numexpr}], which lessen the load on the memory bus.
Cython, a Python to C compiler discussed later in this issue, is
especially useful in cases where code cannot be vectorized easily.

\subsection*{Finite
Differencing\phantomsection\addcontentsline{toc}{subsection}{Finite
Differencing}\label{finite-differencing}}

The derivative on a discrete sequence is often computed using finite
differencing.  Slicing makes this operation trivial.

Suppose we have an \DUrole{math}{n + 1} length vector and perform a
forward divided difference.

\begin{verbatim}
In [36]: x = np.arange(0, 10, 2)
 
In [37]: x
Out[37]: array([ 0,  2,  4,  6,  8])
 
In [38]: y = x**2
 
In [39]: y
Out[39]: array([  0,   4,  16,  36,  64])
 
In [40]: dy_dx = (y[1:]-y[:-1])/(x[1:]-x[:-1])
 
In [41]: dy_dx
Out[41]: array([ 2,  6, 10, 14, 18])
\end{verbatim}

In this example, \texttt{y{[}1:{]}} takes a slice of the \texttt{y} array starting
at index 1 and continuing to the end of the array. \texttt{y{[}:-1{]}} takes a
slice which starts at index 0 and \protect\TLSdel{contains all elements apart from} \protect\TLSins{continues to one index short of} the
\protect\TLSdel{last.}
\protect\TLSins{end of the array.} Thus \texttt{y{[}1:{]} - y{[}:-1{]}} has the effect of
subtracting, from each element in the array, the element directly
preceding it. Performing the same differencing on the \texttt{x} array and
dividing the two resulting arrays yields the forward divided
difference.

If we assume that the vectors are length \DUrole{math}{n + 2}, then
calculating the central divided difference is simply a matter of
modifying the slices:
\begin{verbatim}
In [42]: dy_dx_c = (y[2:]-y[:-2])/(x[2:]-x[:-2])
 
In [43]: dy_dx_c
Out[43]: array([ 4,  8, 12, 16])
\end{verbatim}

In Pure Python, these operation would be written using a for loop. For
\texttt{x} containing 1000 elements, the NumPy implementation is 100 times
faster.

\subsection*{Creating a grid using broadcasting\phantomsection\addcontentsline{toc}{subsection}{Creating a grid using broadcasting}\label{creating-a-grid-using-broadcasting}}

Suppose we want to produce a three-dimensional grid of distances
\DUrole{math}{$R_{ijk} = \sqrt{i^2 + j^2 + k^2}$} with
\DUrole{math}{$i=-100 \ldots 99$},
\DUrole{math}{$j=-100 \ldots 99$}, and
\DUrole{math}{$k=-100 \ldots 99$}.  In most
vectorized programming languages, this would require forming three
intermediate \DUrole{math}{$200 \times \protect\TLSdel{200 \times} 200$} arrays, \DUroletitlereference{i}, \DUroletitlereference{j}, and \DUroletitlereference{k} as in:
\begin{verbatim}
In [44]: i, j, k = np.mgrid[-100:100, -100:100, 
    ...: -100:100]
 
In [45]: print i.shape, j.shape, k.shape
(200, 200, 200) (200, 200, 200) (200, 200, 200)
 
In [46]: R = np.sqrt(i**2 + j**2 + k**2)
 
In [47]: R.shape
Out[47]: (200, 200, 200)
\end{verbatim}

Note the use of the special \texttt{mgrid} object, which produces a
meshgrid when sliced.

In this \protect\TLSdel{example} \protect\TLSins{case} we have allocated 4 named arrays, i, j, k, R and an
additional 5 temporary arrays over the course of the operation. Each
of these arrays contains roughly 64MB of data resulting in a total
memory allocation of \textasciitilde{}576MB.  In total, 48 million operations are
performed: \DUrole{math}{$200^3$} to square each array and
\DUrole{math}{$200^3$} per
addition.

In a non-vectorized language, no temporary arrays need to be allocated
when the output values are calculated in a nested for-loop, e.g. (in
C):
\begin{verbatim}
int R[200][200][200];
int i, j, k;
 
for (i = -100; i < 100; i++)
  for (j = -100; j < 100; j++)
    for (k = -100; k < 100; k++)
      R[i][j][k] = sqrt(i*i + j*j + k*k);
\end{verbatim}
\begin{figure}[t]
\begin{DUlegend}
\hspace*{-.05\linewidth}%
\includegraphics[width=0.5\linewidth]{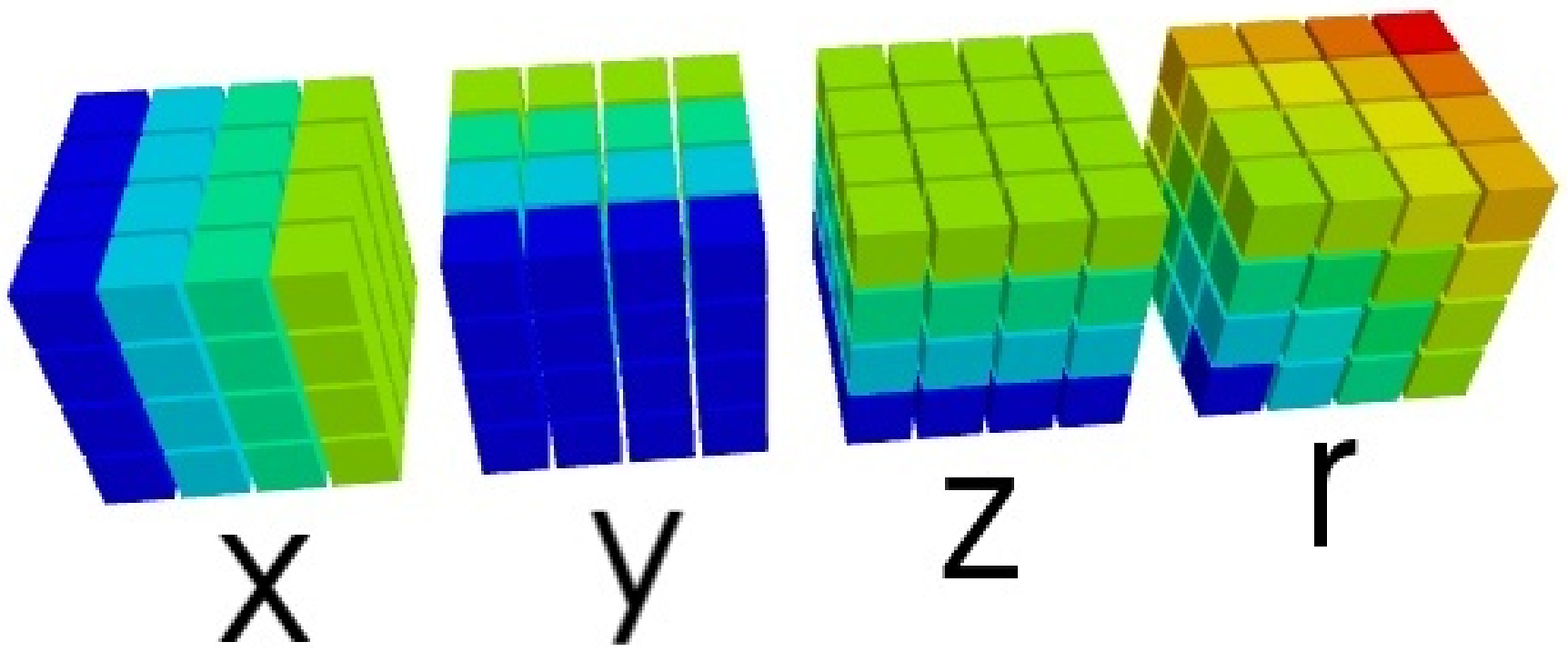}
\hfill%
\includegraphics[width=0.5\linewidth]{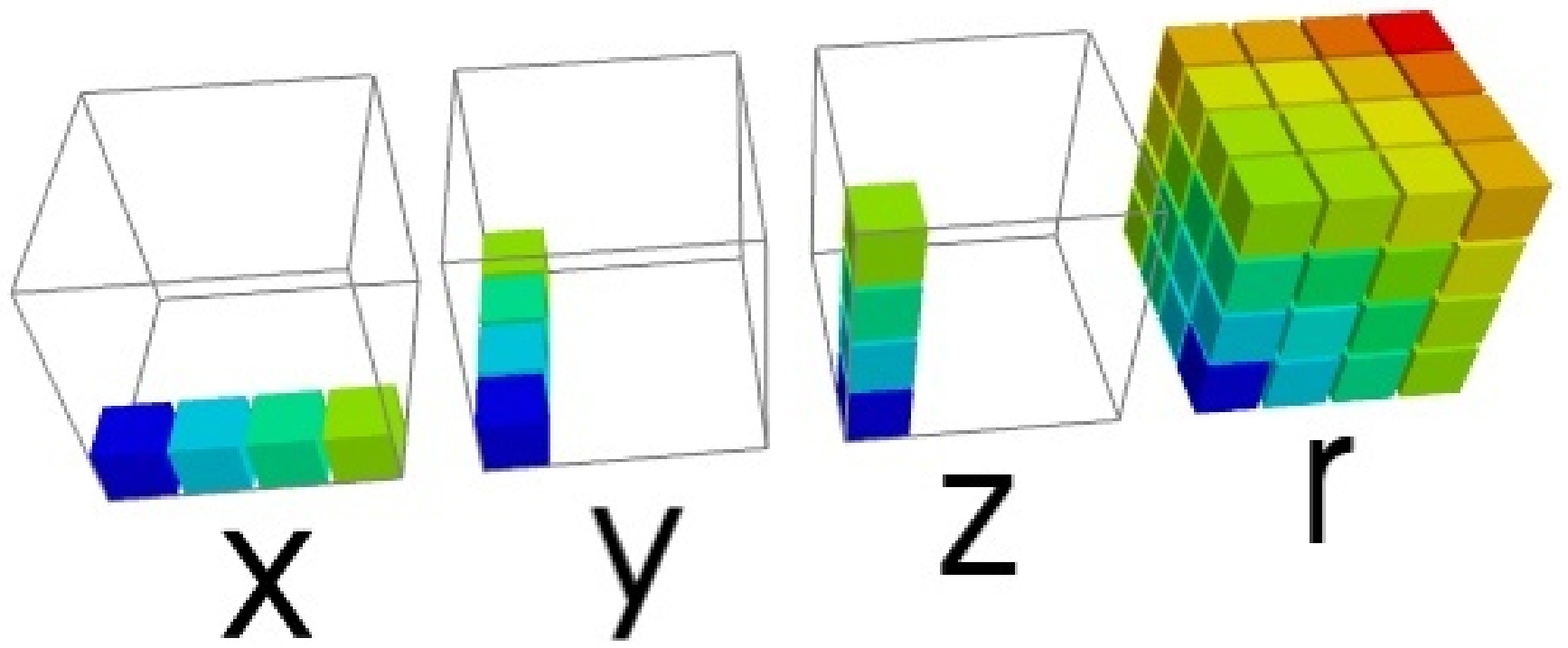}
\hspace*{-.05\linewidth}%
\end{DUlegend}
\caption{Computing dense grid values without and with broadcasting. Note
how, with broadcasting, much less memory is used.
}
\end{figure}

We can achieve a similar effect using NumPy's broadcasting facilities.
Instead of constructing large temporary arrays, we instruct NumPy to
combine three one-dimensional vectors (a row-vector, a column-vector
and a depth-vector) to form the three-dimensional result.
Broadcasting does not require large intermediate arrays.

First, construct the three coordinate vectors (\texttt{i} for the
x-axis, \texttt{j} for the y-axis and \texttt{k} for the z-axis):
\begin{verbatim}
# Construct the row vector: from -100 to +100
i = np.arange(-100, 100).reshape(200, 1, 1)
\end{verbatim}
 
\begin{verbatim}
# Construct the column vector
j = np.reshape(i, (1, 200, 1))
 
# Construct the depth vector
k = np.reshape(i, (1, 1, 200))
\end{verbatim}

NumPy also provides a short-hand for the above construction, namely

\begin{verbatim}
i, j, k = np.ogrid[-100:100, -100:100, -100:100]
\end{verbatim}

Note how the arrays contain the same number of elements, but that they
have different orientations.  We now let NumPy broadcast \texttt{i}, \texttt{j}
and \texttt{k} to form the three-dimensional result, as shown in Fig. 1.:

\begin{verbatim}
In [48]: R = np.sqrt(i**2 + j**2 + k**2)
 
In [49]: R.shape
Out[49]: (200, 200, 200)
\end{verbatim}

Here, the total memory allocation is only 128MB: 4 named arrays
totalling \textasciitilde{}64Mb (1.6KB * 3 + 64MB) and 5 temporary arrays of \textasciitilde{}64MB
(1.6KB * 3 + 320KB + 64MB). A total of approximately 16 million
operations are performed: 200 to square each array, \DUrole{math}{$200^2$} for
the first addition, and \DUrole{math}{$200^3$} each for the second addition as
well as for the square root.

When using naive vectorization, calculating \texttt{R} requires 410ms to
compute.  Broadcasting reduces this time to 182ms--a factor 2
speed-up along with a significant reduction in memory use.

\subsection*{Computer Vision\phantomsection\addcontentsline{toc}{subsection}{Computer Vision}\label{computer-vision}}

Consider an \DUrole{math}{$n \times 3$} array of \protect\TLSdel{three-dimensional point
coordinates} \protect\TLSins{three dimensional points} and a
\DUrole{math}{$3 \times 3$} camera matrix:

\begin{verbatim}
points = np.random.random((100000, 3))
camera = np.array([[500.,   0., 320.],
                   [  0., 500., 240.],
                   [  0.,   0.,   1.]])
\end{verbatim}

Often, we want to transform the 3D coordinates into their 2D pixel
locations on the image, as viewed by the camera.  This operation
involves taking the matrix dot product of each point with the camera
matrix, and then dividing the resulting vector by its third
component. With NumPy, it is written as:

\smallskip
\begin{verbatim}
# Perform the matrix product on the coordinates
vecs = camera.dot(points.T).T
 
# Divide resulting coordinates by their z-value
pixel_coords = vecs/vecs[:, 2, np.newaxis]
\end{verbatim}

The \texttt{dot} function\footnote{
The \texttt{dot} function leverages accelerated BLAS
implementations, if available.
}
implements the matrix product, in contrast to
the element-wise product \texttt{*}.  It can be applied to one- or
two-dimensional arrays.

This code executes in 9 miliseconds--a 70x speedup over a Python
for-loop version.

Aside from the optimized NumPy dot product, we make use of NumPy's
array operations with element-by-element division and the broadcasting
machinery. The code
\texttt{new\_vecs / new\_vecs{[}:, 2, np.newaxis{]}} divides each column of
\texttt{new\_vecs} by its third column
(in other words, each row is divided by its third element). The
\texttt{np.newaxis} index is used to \protect\TLSdel{insert a new axis, thereby changing} \protect\TLSins{change} \texttt{new\_vecs{[}:, 2{]}} into a
column-vector so that broadcasting may take place.

\DUtransition

The above examples show how vectorization provides a powerful and
efficient means of operating on large arrays, without compromising
clear and concise code or relinquishing control over aspects such as
memory allocation.

It should be noted that vectorization and broadcasting is no panacea;
for example, when repeated operations take place on very large chunks
of memory, it may be better to use an outer for-loop combined with a
vectorised inner loop to make optimal use of the system cache.

\section*{Sharing data\phantomsection\addcontentsline{toc}{section}{Sharing data}\label{sharing-data}}

As shown above, performance is often improved by preventing repeated
copying of data in memory. In this section, we show how NumPy may make
use of foreign memory--in other words, memory that is not allocated or
controlled by NumPy--without copying data.

\subsection*{Efficient I/O with memory mapping\phantomsection\addcontentsline{toc}{subsection}{Efficient I/O with memory mapping}\label{efficient-i-o-with-memory-mapping}}

An array stored on disk may be addressed directly without copying it
to memory in its entirety.  This technique, known as \emph{memory mapping},
is useful for addressing only a small portion of a very large \protect\TLSdel{array
(such as produced by an external instrument, for example).} \protect\TLSins{array.}
NumPy supports memory mapped arrays with the same interface as any
other NumPy array.  First, let us construct such an array and fill it
with some data:
\begin{verbatim}
In [50]: a = np.memmap('/tmp/myarray.memmap', 
    ...: mode='write', shape=(300, 300), 
    ...: dtype=np.int)

# Pretend "a" is a one-dimensional, 300*300 
# array and assign values into it
In [51]: a.flat = np.arange(300* 300)
 
In [52]: a
Out[52]:
memmap([[    0,     1, ...,   298,   299],
        [  300,   301, ...,   598,   599],
        [  600,   601, ...,   898,   899],
        ...,
        [89100, 89101, ..., 89398, 89399],
        [89400, 89401, ..., 89698, 89699],
        [89700, 89701, ..., 89998, 89999]])
\end{verbatim}

When the ``flush'' method is called, its data is written to disk:

\begin{verbatim}
In [53]: a.flush()
\end{verbatim}

The array can now be loaded and parts of it manipulated; calling
``flush'' writes the altered data back to disk:

\smallskip
\begin{verbatim}
# Load the memory mapped array
In [54]: b = np.memmap('/tmp/myarray.memmap',
    ...: mode='r+', shape=(300, 300), 
    ...: dtype=np.int)
 
# Perform some operation on the elements of b
In [55]: b[100, :] *= 2
 
# Store the modifications to disk
In [56]: b.flush()
\end{verbatim}

\subsection*{The \emph{array interface} for foreign blocks of memory\phantomsection\addcontentsline{toc}{subsection}{The array interface for foreign blocks of memory}\label{the-array-interface-for-foreign-blocks-of-memory}}

Often, NumPy arrays have to be created from memory constructed and
populated by foreign code, e.g., a result produced by an external C++
or Fortran library.

To facilitate such exchanges without copying the already allocated
memory, NumPy defines an \textbf{array interface} that specifies how a
given object exposes a block of memory.  NumPy knows how to view any
object with a valid \texttt{\_\_array\_interface\_\_} dictionary attribute as an
array.  Its most important values are \texttt{data} (address of the data in
memory), \texttt{shape} and \texttt{typestr} (the kind of elements stored).

The following example defines a \texttt{MutableString} class that allocates
a string \texttt{\_s}.  The MutableString represents a foreign block of
memory, now made made available to NumPy by defining the
\texttt{\_\_array\_interface\_\_} dictionary.

Note the use of the ctypes library, which allows Python to execute
code directly from dynamic C libraries.  In this instance, we use its
utility function, \texttt{create\_string\_buffer}, to allocate a string in
memory and \texttt{addressof} to establish its position in memory.

\begin{verbatim}
import ctypes
 
class MutableString(object):
  def __init__(self, s):
    # Allocate string memory
    self._s = ctypes.create_string_buffer(s)

    self.__array_interface__ = {
      # Shape of the array
      'shape': (len(s),),

      # Address of data,
      # the memory is not read-only
      'data': (ctypes.addressof(self._s), False),

      # Stores 1-byte unsigned integers.
      # "|" indicates that Endianess is
      # irrelevant for this data-type.
      'typestr': '|u1',
      }

  def __str__(self):
    "Convert to a string for printing."
    return str(buffer(self._s))
\end{verbatim}

\medskip
The above class is instantiated, after which NumPy is asked to
interpret it as an array, which is possible because of its
\texttt{\_\_array\_interface\_\_} attribute.

\smallskip
\begin{verbatim}
# Create an instance of our mutable string class
In [57]: m = MutableString('abcde')
 
# View the character byte values as an array
In [58]: am = np.asarray(m)
 
In [59]: am
Out[59]: array([ 97,  98,  99, 100, 101], 
 dtype=uint8)

# Modify the values of the array
In [60]: am += 2
 
In [61]: am
Out[61]: array([ 99, 100, 101, 102, 103], 
 dtype=uint8)
\end{verbatim}
 
\begin{verbatim}
# Since the underlying memory was updated, 
# the string now has a different value.
In [62]: print m
cdefg
\end{verbatim}

This example illustrates how NumPy is able to interpret any block of
memory, as long as the necessary information is provided via an
\texttt{\_\_array\_interface\_\_} dictionary.

\subsection*{Structured data-types to expose complex data\phantomsection\addcontentsline{toc}{subsection}{Structured data-types to expose complex data}\label{structured-data-types-to-expose-complex-data}}

NumPy arrays are homogeneous, in other words, each element of the
array has the same data-type.  Traditionally, we think of the
fundamental data-types: integers, floats, etc.  However, NumPy arrays
may also store compound elements, such as the combination \texttt{(1, 0.5)}
-- an \protect\TLSdel{integer} \protect\TLSins{array} \emph{and} a float.  Arrays that store such compound elements
are known as structured arrays.

Imagine an experiment in which measurements of the following fields
are recorded:
\begin{itemize}

\item Timestamp in nanoseconds (a 64-bit unsigned integer)

\item Position (x- and y-coordinates, stored as floating point numbers)

\end{itemize}

We can describe these fields using a single data-type:

\begin{verbatim}
In [63]: dt = np.dtype([('time', np.uint64),
    ...:                ('pos', [('x', float),
    ...:                        ('y', float)])])
\end{verbatim}

An array of measurements can then be constructed using this data-type as

\begin{verbatim}
In [64]: x = np.array([(1, (0, 0.5)), 
    ...:               (2, (0, 10.3)), 
    ...:               (3, (5.5, 1.1))], 
    ...:              dtype=dt)
\end{verbatim}

The individual fields of such a structured array can be queried:

\begin{verbatim}
# Display all time-stamps
In [65]: x['time']
Out[65]: array([1, 2, 3], dtype=uint64)
 
# Display x-coordinates for all timestamps >= 2
In [66]: times = (x['time'] >= 2)
 
In [67]: print times
[False  True  True]
 
In [68]: x[times]['pos']['x']
Out[68]: array([ 0. ,  5.5])
\end{verbatim}

\protect\TLSdel{The above example also shows how arrays are indexed using boolean
``masks''.  For example, when \texttt{x} is indexed by \texttt{times}, only the
elements corresponding to a \texttt{True} value in \texttt{times} are returned.}

Structured arrays are useful for reading complex binary \protect\TLSdel{files that
contain sequential, fixed-length records.} \protect\TLSins{files.}
Suppose we have a file ``foo.dat'' that contains binary data structured
according to the data-type 'dt' introduced above: for each record, the
first 8 bytes are a 64-bit unsigned integer time stamp and the next 16
bytes a position comprised of a 64-bit floating point 'x' position and
a floating point 'y' position.

Loading such data manually is cumbersome: for each field in the
record, bytes are read from file and converted to the appropriate
data-type, taking into consideration factors such as endianess.

In contrast, using a structured data-type simplifies this operation to
a single line of code:

\begin{verbatim}
In [69]: data = np.fromfile("foo.dat", dtype=dt)
\end{verbatim}

\section*{Conclusion\phantomsection\addcontentsline{toc}{section}{Conclusion}\label{conclusion}}

We show that the N-dimensional array introduced by NumPy is a
high-level data structure that facilitates vectorization of for-loops.
Its sophisticated memory description allows a wide variety of
operations to be performed without copying any data in memory,
bringing significant performance gains as data-sets grow large. Arrays
of multiple dimensions may be combined using \emph{broadcasting} to reduce
the number of operations performed during numerical computation.

When NumPy is skillfully applied, most computation time is spent on
vectorized array operations, instead of in Python for-loops (which are
often a bottleneck).  Further speed improvements are achieved \protect\TLSdel{using} \protect\TLSins{by}
optimizing compilers, such as Cython, which \protect\TLSdel{allow} better \protect\TLSdel{control over} \protect\TLSins{exploit} cache
effects.

\protect\TLSdel{In addition to low-level array operations, NumPy provides sub-packages
for linear algebra, FFTs, random number generation and polynomial
manipulation.  Larger scientific packages, such as SciPy, are, in
turn, built on this infrastructure.}

NumPy and similar projects foster an environment in which numerical
problems may by described using high-level code, thereby opening the
door to scientific code that is both transparent and easy to maintain.

\section*{References\phantomsection\addcontentsline{toc}{section}{References}\label{references}}
\begin{description}
\item[\hypertarget{ipython}{}[IPython{]}]
Fernando Perez, Brian E. Granger. IPython: A System for
Interactive Scientific Computing, Computing in Science and
Engineering, vol. 9, no. 3, pp. 21-29, May/June 2007,
doi:10.1109/MCSE.2007.53.

\item[\hypertarget{cython}{}[cython{]}]
S. Behnel, R. Bradshaw, D. S Seljebotn, G. Ewing et
al. C-Extensions for Python. \href{http://www.cython.org}{http://www.cython.org}.

\item[\hypertarget{theano}{}[theano{]}]
J. Bergstra. Optimized Symbolic Expressions and GPU
Metaprogramming with Theano, Proceedings of the 9th Python in
Science Conference (SciPy2010), Austin, Texas, June 2010.

\item[\hypertarget{numexpr}{}[numexpr{]}]
D. Cooke, F. Alted, T. Hochberg, G. Thalhammer, \emph{numexpr}
\href{http://code.google.com/p/numexpr/}{http://code.google.com/p/numexpr/}

\end{description}

\vfill
\framebox{
\begin{minipage}{\linewidth}
NumPy's documentation is maintained using a \protect\TLSdel{Wikipedia-like} \protect\TLSins{WikiPedia-like}
community forum, available at \texttt{http://docs.scipy.org/}.
Discussions take place on the project mailing list
(\texttt{http://www.scipy.org/Mailing\_Lists}).  NumPy is a volunteer
effort.
\end{minipage}
}
\vfill

\end{document}